\documentclass{article}
\usepackage{spconf,amsmath,amssymb,graphicx}
\usepackage{booktabs}
\usepackage{multirow}
\usepackage{array}
\usepackage{xcolor}
\usepackage{colortbl}
\usepackage{tikz}
\usetikzlibrary{positioning,fit,backgrounds,arrows.meta,calc}
\usepackage{balance}
\usepackage{url}
\newcommand{\unnumberedsection}[1]{%
  \par\vspace{1.2em}%
  {\centering\bf\uppercase{#1}\par}%
  \vspace{0.6em}%
  \noindent}
\title{Timestamped Hindi speech transcription using Whisper}

\name{Sanat Kumar Agrawal and Srikanth Raj Chetupalli}
\address{Electrical Engineering, Indian Institute of Technology Bombay, Mumbai, India \\
         \texttt{\{sanat.agrawal, srikanthrajch\}@iitb.ac.in}}

\begin{document}
\ninept
\maketitle

\begin{abstract}
Time accurate speech transcription is essential in critical applications, such as medical conversation transcription, reading assessment, and atypical speech recognition. Encoder-decoder automatic speech recognition models, such as Whisper, do not produce word-level timestamps out-of-the-box for all languages. Existing approaches to word-level time stamp estimation either use an external CTC model capable of frame level phoneme/character prediction which is computationally expensive, or use the cross attention scores internal to the model which gives coarse time stamps. 
We consider timestamped speech transcription for Hindi in this paper. We propose a CTC-based approach, in which we train a low-complexity character level CTC head directly on the Whisper encoder output, to address the complexity and granularity limitations of existing works. We evaluate the system using manually-annotated word-level timestamps. Experiments show that the proposed internal-CTC based approach performs similar to using an external CTC model and better than the cross attention score-based approaches. Further, we study the usage of CTC-head output to detect and mitigate hallucinations in conversational speech. Our results show that a simple token length criterion derived from the CTC output can improve the transcription quality for short segments where hallucinations are more probable.

\end{abstract}


\section{Introduction}
\label{sec:intro}
Large-scale weakly supervised models, such as Whisper \cite{radford2023whisper}, have shown remarkable success and robustness for automatic speech recognition (ASR). However, the generated transcripts often are not grounded in the input utterance, typically noted when the model hallucinates. Timestamped speech transcription aims to predict start and end times at the word/token level and can be used to check for transcription conformity with respect to the speech input. Further, timestamps are needed in applications such as spoken-content retrieval, keyword spotting, computer-assisted language learning, reading and fluency assessment, and automatic subtitling. This paper addresses timestamped speech transcription for low-resource settings, in particular for Hindi, using Whisper \cite{radford2023whisper}.


\noindent Existing approaches to timestamped speech transcription using Whisper can be grouped into two categories. The first approach uses dynamic time warping (DTW) on the decoder-encoder cross-attention scores 
to derive timestamp information \cite{lintoai2023whispertimestamped, whisper-dtw, wagner2024crisperwhisper}. The method was initially developed in \cite{lintoai2023whispertimestamped, whisper-dtw} and CrisperWhisper \cite{wagner2024crisperwhisper} improved the timestamp accuracy by carefully adjusting the tokenizer and extending it for verbatim transcription. The DTW approach of \cite{lintoai2023whispertimestamped,whisper-dtw} is scalable and works on all variants (monolingual, multilingual) of Whisper, but its timestamp error is typically higher than the second approach based on connectionist temporal classification (CTC) forced alignment.
WhisperX \cite{bain2023whisperx} uses CTC-based wav2vec2.0 model \cite{baevski2020wav2vec2}, trained for character/phoneme level recognition, to force align the Whisper transcripts and derive the timestamps. Framelevel posterior available in CTC approach enables fine timestamp estimation. 
However, the use of a second model, which is also language specific in \cite{bain2023whisperx}, limits the capabilities of the framework while increasing the complexity. Further, the timestamp accuracy depends jointly on the performance of Whisper and wav2vec2.0 models, which is not uniform across different acoustic environments. Other similar approaches, using HMM-based forced alignment \cite{mcauliffe2017mfa} and massively multilingual aligners \cite{pratap2024mms} also need a second model and suffer from drawbacks similar to WhisperX. Both CrisperWhisper~\cite{wagner2024crisperwhisper} and WhisperX~\cite{bain2023whisperx} were developed for English and to the best of our knowledge, models capable of timestamped transcription are not available for Hindi, and more generally for Indic languages.

\noindent We consider timestamped speech transcription for Hindi in this paper. Our approach builds on 
WhisperX~\cite{bain2023whisperx}, but instead of an external CTC-based model, we propose to train a light-weight CTC head on the Whisper encoder output. The approach is parameter and compute efficient as it does not need a second model, and our experiments show that the performance is comparable to that of WhisperX~\cite{bain2023whisperx}. The idea of a CTC head on Whisper encoder output is first proposed in \cite{zhou2025streamingwhisper} in the context of streaming transcription, and later used by \cite{polok2025dicow} to mitigate hallucination. Both, \cite{zhou2025streamingwhisper} and \cite{polok2025dicow}, use variants of joint CTC-attention decoding \cite{watanabe2017hybrid} and the CTC head was trained to predict a large subset of Whisper output tokens ($8$k in \cite{zhou2025streamingwhisper} and $\approx 50$k tokens in \cite{polok2025dicow}). In the present work, we use character-level CTC head, with a vocabulary of size $73$, leading to a small number of additional parameters. We also adapted CrisperWhisper~\cite{wagner2024crisperwhisper} to Hindi and compared its performance with the proposed approach. 

\noindent One of the main issues with Whisper for transcription is hallucination, especially in short speech segments. We investigate the use of the CTC head to detect and correct the hallucinated transcriptions in conversational settings. We show that simple rules based on the expected transcription length derived from CTC output are sufficient for this task. A demonstration of the proposed system, along with the models, code, and the manual annotation data are available online\footnote{\texttt{\scriptsize{https://sanat-agrwl.github.io/hindi-word-boundaries-demo}}}.

\vspace{-3pt}
\section{Timestamped speech transcription}
\vspace{-2pt}
\label{sec:method}
\begin{figure}[t]
\centering
\resizebox{0.98\columnwidth}{!}{%
\begin{tikzpicture}[
  every node/.style={font=\small},
  box/.style={draw, rounded corners=3pt, minimum height=6mm, align=center},
  enc/.style={box, text width=15mm, fill=teal!10, draw=teal!55!black,align=center},
  dec/.style={box, text width=15mm, fill=gray!10, draw=gray!50,align=center},
  head/.style={box, text width=15mm, fill=violet!14, draw=violet!60!black, thick,align=center},
  fa/.style={box, text width=15mm, fill=violet!14, draw=violet!60!black, thick,align=center},
  ext/.style={box, minimum width=26mm, fill=orange!10, draw=orange!65!black, dashed},
  ar/.style={-{Latex[length=1.8mm,width=1.4mm]}, line width=0.7pt},
]
\node[] (aud) at (0,0) {Audio};
\node[enc] (enc) at (20mm,0) {Whisper\\encoder };
\node[dec] (dec) at (48mm, 6mm) {Whisper decoder};
\node[head] (hd)  at (48mm,-7mm) {Linear head-CTC};
\node[] (txt) at (70mm, 6mm) {Transcript};
\node[fa] (fa) at (70mm, -7mm) {Forced alignment};

\node[text width=15mm, align=center] (ts) at (95mm,-7mm)
     {word timestamps};
\draw[ar] (aud) -- (enc);
\draw[ar] (enc.east) -- ++(2mm,0) |- (dec.west);
\draw[ar] (enc.east) -- ++(2mm,0) |- (hd.west);
\draw[ar] (dec) -- (txt);
\draw[ar] (hd)  -- (fa);
\draw[ar] (fa)  -- (ts);
\draw[ar, dashed] (txt.south) -- ++(0,-6mm) -| (fa.north);

\node[] (aud2) at (0,-25mm) {Audio};
\node[enc] (enc2) at (20mm,-25mm) {Whisper\\encoder };
\node[dec] (dec2) at (48mm, -25mm) {Whisper decoder};
\node[] (txt2) at (70mm, -25mm) {Transcript};
\draw[ar] (aud2) -- (enc2);
\draw[ar] (enc2) -- (dec2);
\draw[ar] (dec2) -- (txt2);
\node[ext] (w2v) at (35mm,-38mm) {CTC wav2vec2.0};
\node[fa] (fa2) at (70mm, -38mm) {Forced alignment};
\draw[ar] (txt2) -- (fa2);
\draw[ar, black] (aud2.south) |- (w2v.west);
\draw[ar, black] (w2v.east) -- (fa2);
\node[text width=15mm, align=center] (ts2) at (95mm,-38mm)
     {word timestamps};
\draw[ar] (fa2) -- (ts2);
\begin{scope}[on background layer]
  \node[draw=gray!35, rounded corners=5pt, fill=gray!3, inner xsep=3mm, inner ysep=3mm,
        fit=(aud)(enc)(dec)(hd)(txt)(ts)] (boxtop) {};
  \node[draw=gray!35, rounded corners=5pt, fill=gray!3, inner xsep=3mm, inner ysep=3mm,
        fit=(aud2)(enc2)(dec2)(txt2)(w2v)(fa2)(ts2)] (boxbot) {};
\end{scope}
\node[anchor=north east, font=\small\bfseries, text=gray!50!black]
     at ([shift={(-1.5mm,-1.0mm)}]boxtop.north east) {Proposed};
\node[anchor=north east, font=\small\bfseries, text=gray!50!black]
     at ([shift={(-1.5mm,-1.0mm)}]boxbot.north east) {WhisperX};
\end{tikzpicture}}
\vspace{-5pt}
\caption{Block schematic of the proposed CTC-based approach (top) and WhisperX \cite{bain2023whisperx} (bottom). 
}
\label{fig:arch}
\vspace{-5pt}
\end{figure}
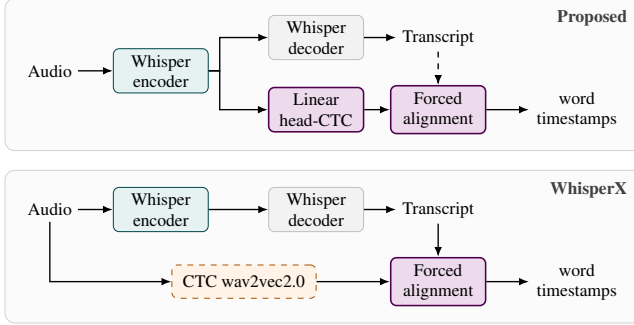

Figure~\ref{fig:arch} (top) shows a block schematic of the proposed approach. Whisper is an encoder-decoder model, with it's encoder producing fine frame-level representations at a $50$~Hz rate. We propose to train a CTC head, which is a single linear layer, on the encoder output to predict posteriors on a chosen vocabulary. This is in contrast to WhisperX~\cite{bain2023whisperx}, shown in Figure~\ref{fig:arch} (bottom), where an external wav2vec2.0~\cite{baevski2020wav2vec2} model trained using CTC is used for frame level posterior computation. We chose a character-level vocabulary of size $73$, composed of $71$ Devanagari characters, an explicit space symbol and the CTC blank symbol in this work. The \emph{character} vocabulary is chosen because the alignment resolution depends on the granularity of the units emitted by the CTC head. Devanagari byte pair encoding (BPE) tokens, used by Whisper, span only $0.77$ characters on average on our test transcripts.
Hence, the predicted token boundaries often do not coincide with the character boundaries and they are irregular for boundary placement. The compact head, due to the small vocabulary also leads to computationally efficient training and timestamp estimation during inference.

\noindent Word timestamps are obtained by using forced alignment \cite{kurzinger2020ctcseg, graves2006ctc} between the frame posteriors and the transcript words predicted by the Whisper decoder. Forced alignment uses DTW to fit a monotonic mapping between the transcript words and the frame-level character posteriors. A word spans a contiguous run of characters, hence we select the first character's frame index as the start and the last character's frame index as the end to compute the time stamps.

\vspace{-5pt}
\section{Experimental Setup}
\label{sec:setup}

\noindent\textbf{Architecture and models.}
We used Whisper large-v2 \cite{radford2023whisper} architecture in all our experiments with pretrained weights from a model finetuned for Hindi\footnote{{\texttt{hf.co/vasista22/whisper-hindi-large-v2}}}, which we refer to as Hindi-Whisper in the rest of the paper. The encoder in the model generates $1280$ dimensional representations every $20$~ms for a fixed $30$~s segment. In the proposed CTC-based approach, we used a vocabulary of $73$ symbols, which includes $71$ Hindi characters derived from the IndicVoices-R reference transcripts, an explicit space symbol, and the blank symbol. The input, output dimensions of
the linear layer were $1280$ and $73$, respectively, leading to $\approx 93.5$K additional parameters. 

\begin{table}[t]
\centering
\setlength{\tabcolsep}{5pt}
\caption{Model configurations investigated in this work.  Stage~1 corresponds to an optional cross-entropy fine-tuning of the
Whisper-Hindi base on IndicVoices-R and stage~2 trains the CTC head and, optionally, Whisper's encoder and decoder blocks.  
}
\label{tab:config}
\vspace{3pt}
\begin{tabular}{llccc}
\toprule
 & Stage~1 & \multicolumn{3}{c}{Stage~2: trained} \\
\cmidrule(lr){3-5}
Config & CE fine-tune & CTC head & Encoder & Decoder \\
\midrule
M1 &            & \checkmark &            &            \\
M2 &            & \checkmark & \checkmark &            \\
M3 &            & \checkmark & \checkmark & \checkmark \\
\midrule
M4 & \checkmark & \checkmark &            &            \\
M5 & \checkmark & \checkmark & \checkmark &            \\
M6 & \checkmark & \checkmark & \checkmark & \checkmark \\
\bottomrule
\end{tabular}
\vspace{-10pt}
\end{table}

We developed six different models (M1-M6) corresponding to the configurations shown in Table~\ref{tab:config}. Configurations M1-M3 start from the pretrained Hindi-Whisper model. In M1, only the proposed head was trained with the CTC objective. Hence, the transcription quality of M1 will be the same as that of the Hindi-Whisper model, but the added CTC head facilitates word timestamp computation. In M2, Whisper encoder was fine tuned along with the head with the same objective, and in M3, both the encoder and decoder were trained. For training M3, we used the hybrid CTC/attention objective $\mathcal{L}=\alpha\mathcal{L}_{\mathrm{CTC}}+(1{-}\alpha)\mathcal{L}_{\mathrm{CE}}$~\cite{watanabe2017hybrid}, where $\mathcal{L}_{\mathrm{CE}}$ is the cross-entropy loss computed over the BPE output space of Whisper decoder. Training for configurations M4-M6, resembles that of M1-M3, respectively. However, the model training starts from the Hindi-Whisper variant, which was initially finetuned on our training data.



\noindent\textbf{Datasets.}
Experiments were carried out using the Hindi subset of IndicVoices-R \cite{indicvoicesr2024}, 
having $26.3$K training utterances. The corresponding test set is a subset of the IndicVoices-R test split. The test set has $125$ utterances comprising $1.89$K reference words, obtained by keeping the utterances that the external wav2vec2.0 aligner could align with a minimum per-word confidence of $0.30$ (median $0.45$).

\noindent Further, we evaluated on three additional corpora, (i) MPS-Hindi, (ii) GramVaani, and (iii) DISPLACE-M, covering out of domain, noisy, and conversational scenarios. MPS-Hindi is a children's oral-reading corpus collected similar to the protocol described in \cite{gothi2024miscue, raman2025oralreading}. The dataset has $3.6$K recordings with  ${\approx}240$K reference words in total. 
GramVaani \cite{bhanushali2022gramvaani} is spontaneous telephone speech in regional Hindi ($1{,}032$ recordings, ${\approx}27$K words), and DISPLACE-M \cite{dhanya2026displacem} is the Hindi development set of the Third DISPLACE Challenge.
The dataset contains conversations between frontline health workers and care seekers in rural India, with spontaneous dialogue, foreground overlap, background speech and
dialectal variation ($78$ recordings, $163.8$K words). Among the three, only GramVaani corpus was part of the training set of Hindi-Whisper, and our evaluation on MPS-Hindi and DISPLACE-M represents true generalization. 


\noindent\textbf{Training.}
We trained the proposed CTC-based models incrementally. First, the configurations M1 and M4 were trained; that is, only the CTC head of the model was trained for $15$~epochs using AdamW optimizer \cite{loshchilov2019adamw} with Cosine annealing learning rate schedule \cite{loshchilov2017sgdr} from $3\times10^{-4}$ to $3\times10^{-6}$ and an effective batchsize of $32$. Then, Whisper's encoder is also finetuned for additional $3$ epochs alongside the CTC head for M2 and M5 configurations. Similarly, training for M3 and M6 starts from M1 and M4 models, respectively, and both the encoder and decoder were finetuned for $3$ epochs in these two configurations. 
We used different learning rates for the encoder-decoder pair ($10^{-5}$) and the CTC-head ($10^{-4}$), along with cosine annealing. CTC loss weight $\alpha$ was chosen to be $0.1$ for the M3 model and $0.3$ for M6 and was fixed based on limited initial hyperparameter search. Since Whisper training pads all training examples to $30$~s duration, we discarded the frames corresponding to the padded samples while training the CTC head. 


\noindent\textbf{Baseline systems.}
Existing ASR pipelines for Hindi do not provide timestamped transcriptions. Hence, we adapted the WhisperX~\cite{bain2023whisperx} and CrisperWhisper~\cite{wagner2024crisperwhisper} approaches to Hindi for a fair evaluation. For WhisperX adaptation, we used the version of wav2vec2.0\footnote{\scriptsize\texttt{hf.co/theainerd/Wav2Vec2-large-xlsr-hindi}}, which is an XLSR-53 \cite{conneau2021xlsr} model fine-tuned for Hindi as the external CTC model and used Hindi-Whisper model for transcription. For CrisperWhisper adaptation, we followed the steps proposed in \cite{wagner2024crisperwhisper}, that is, (i) constructed a new tokenizer ($\approx 50.4$K entries) by detaching the space ("\_") token from all tokens, (ii) finetuned the Hindi-Whisper model with the modified tokenizer, and (iii) re-estimated the $15$ attention heads used for DTW on a calibration set of $300$ utterances. The time stamps were additionally compared with wav2vec2.0 and Nemo forced aligner (NFA) \cite{NemoFA}. For NFA, we used the CTC-based IndicConformer model\footnote{\scriptsize\texttt{hf.co/ai4bharat/indicconformer\_stt\_hi\_hybrid\_ctc\_rnnt\_large}} to compute frame level posteriors.

\noindent\textbf{Inference.} ASR inference uses the HuggingFace \texttt{pipeline} of Hindi-Whisper for every system with \texttt{language="hi"} and
\texttt{task="transcribe"}. We evaluate the models in both, greedy, and beam search (with a beam size of 5) conditions.


\noindent\textbf{Metrics.}
We report word error rate (WER), character error rate (CER) and word-boundary error as the performance measures. WER was computed using \texttt{jiwer} \cite{vanheeswijk2021jiwer} after standard text normalization. The CER was computed over Devanagari \emph{grapheme clusters}, where each cluster is a base character with its combining marks and conjuncts.
Word-boundary error was computed as the mean absolute difference (in ms) pooled over start times and end times for words matching between the reference and hypothesis transcripts. Longest common subsequence (LCS) algorithm was used to align the two transcripts.

\noindent\textbf{Manual annotation.}  IndicVoices-R does not have word-level timestamp
annotations, and as we show later in Section~\ref{sec:timestamp_only}, existing forced aligners using CTC-based models often have biased estimates for word boundaries. Hence, we manually annotated the $125$-utterance test set of IndicVoices-R.  Annotators were shown the audio and the ground truth transcript, and they were asked to mark the word boundaries. 
$100$ of the reference utterances were annotated twice by two different annotators, and 
the remaining $25$ were annotated once.  Every utterance contributes one human reference to the system comparisons, giving $3{,}784$ scored boundaries.

\vspace{-3pt}
\section{Results}
\vspace{-2pt}
\label{sec:results}
\subsection{Timestamp estimation}
\label{sec:timestamp_only}
First, we assess the quality of the estimated timestamps using the manually annotated test samples. To decouple the effect of ASR performance, we force align the CTC output with the ground truth speech transcripts and compute the word boundary errors. Table~\ref{tab:human} shows the error metrics, averaged over the start and end times, for the different models explored in this paper. First, we note that there is a good inter-annotator agreement in the ground truth timestamps (median boundary error is less than one frame duration for Whisper). We see that force alignment using the wav2vec2.0 CTC model gives the least timestamp estimation error (in column marked \textit{Raw}). Rows M1-M6 indicate performance using the CTC head in proposed models. 
Difference in performance of all configurations of proposed approach to wa2vec2.0 (equivalent to WhisperX) is less than $10$~ms. CTC is known to produce sharp frame level posteriors and often predicting the character earlier than the end timestamp. This introduces a systematic bias in the estimates of start and end-times, as shown in the Table. The column \textit{Calibrated} shows the timestamp errors after correcting for the fixed offset, which shows the proposed approach to be performing similar to the wav2vec2.0 approach. 
Table~\ref{tab:human} also shows the performance of NFA aligner using ground truth transcriptions. We find that the NFA aligner has a higher timestamp error compared to wav2vec2.0 approach, and the performance of proposed models is similar to that of NFA. Hence, in the following, we compared our models against timestamps obtained using force alignment with wav2vec2.0 output and ground truth transcripts. 




\begin{table}[t]
\centering\footnotesize
\setlength{\tabcolsep}{3pt}
\caption{Boundary error against \textbf{manual} annotation (ms) over the full $125$-utterance
test set.  
``Calibrated'' removes each system's own two constant offsets, fitted
on $100$ utterances and applied to the held-out $25$ ($5$-fold, so every value is out-of-sample). 
}\label{tab:human}
\vspace{2pt}
\begin{tabular}{lllrrrr}
\toprule
 &   \multicolumn{2}{c}{\textit{Raw}}&\multicolumn{2}{c}{Bias} & \multicolumn{2}{c}{Calibrated} \\
\cmidrule(lr){2-3}\cmidrule(lr){4-5}\cmidrule(lr){6-7}
System &   mean &med. &onset & offset & mean & med. \\
\midrule
\rowcolor{gray!12}
Inter annotator &   \textbf{27.0} &\textbf{18.0} &--- & --- & --- & --- \\
\midrule
wav2vec2.0 &   \textbf{70.9} &\textbf{54.5} &$+36.4$ & $-92.8$ & 43.4 & 30.3 \\
\midrule
M1       &   73.6 &59.0 &$+36.7$ & $-102.4$ & \textbf{41.3} & \textbf{30.5} \\
M2        &   80.4 &64.0 &$+41.2$ & $-114.6$ & 42.8 & 31.5 \\
M3 &   79.2 &63.0 &$+41.2$ & $-112.0$ & 42.2 & 31.0 \\
\rowcolor{blue!8}
M4 &   73.0 &58.0 &$+37.8$ & $-100.2$ & 42.0 & 31.0 \\
M5 &   80.4 &65.0 &$+44.3$ & $-111.8$ & 43.1 & 32.0 \\
M6 &   78.6 &63.0 &$+43.2$ & $-108.8$ & 42.8 & 31.7 \\
\midrule
NFA& 79.5 & 62.0 & $+41.2$ & $-102.2$ & 52.1 & 39.6 \\
\bottomrule
\end{tabular}
\vspace{-15pt}
\end{table}

\subsection{Joint ASR and timestamp estimation performance}
\label{sec:ts}

\begin{table}[t]
\centering\small
\setlength{\tabcolsep}{3.5pt}
\caption{ASR and word-boundary estimation performance on the IndicVoices-R test set.
$N_w$ is the number of LCS-matched word pairs scored. $p_{90}$ is the $90$th percentile of the absolute boundary error and ${\le}50$ the percentage of boundaries within $50$~ms.
'CW' stands for CrisperWhisper. 
}
\vspace{2pt}
\label{tab:wer_wbe}
\begin{tabular}{lrrrrrrrr}
\toprule
 & \multicolumn{2}{c}{ASR~(\%)} & & \multicolumn{4}{c}{vs.\ wav2vec2.0} &{Manual}\\
\cmidrule(lr){2-3}\cmidrule(lr){5-8}
Conf. & WER& CER& $N_w$ & Mean & Med. & $p_{90}$ & ${\le}50$ & {Mean} \\
\midrule
M1                      & 22.9 & 13.4 & 1466 & 20.7 & 13.7 & 39.9 & 93.1  &73.3 \\
M2                      & 23.5 & 14.5 & 1450 & 26.2 & 14.8 & 42.7 & 92.4  & 84.4 \\
\rowcolor{blue!8}
M3    & \textbf{14.4} & \textbf{6.8} & 1629 & 21.5 & 13.5 & 40.7 & 93.0  & 78.8 \\
M4               & 14.6 & 7.0  & 1621 & 20.8 & 13.5 & \textbf{39.5} & 93.0  & \textbf{73.2} \\
M5              & 15.4 & 7.7  & 1606 & 22.8 & 15.1 & 42.0 & 92.7  & 81.3 \\
M6           & 14.9 & 7.2 & 1618 & \textbf{20.7} & 13.7 & 39.8 & \textbf{93.4}  & 79.3 \\
\midrule

CW & 14.7 & 7.1 & 1623 & 107.2 & 60.1 & 236.9 & 40.4  & 87.4 \\
\bottomrule
\end{tabular}
\vspace{-10pt}
\end{table}

Table~\ref{tab:wer_wbe} shows a comparison of ASR and timestamp estimation performance for the transcript predicted by the model. Note that our word boundary comparison is with respect to wav2vec2.0 boundaries computed for the ground truth transcript. We see that, the median deviation from the reference is less than one Whisper frame shift for all the configurations of the proposed model and $<50$~ms for more than $92\%$ of frames. ASR performance of WhisperX is the same as that of M1 system and it's timestamp estimation aligns with the same wav2vec2.0 as used reference in Table~\ref{tab:wer_wbe}. Against manual annotation, the timestamp error of WhisperX, with system generated transcript, is found to be $71.0$~ms.
CrisperWhisper, in contrast, deviates significantly
from the reference. The performance trends are also consistent with the results obtained using ground truth transcripts for force alignment (Table~\ref{tab:human}). 

\noindent From the ASR performance, we find that, a significant WER improvement is obtained by either finetuning both encoder and decoder along side the CTC head (as in M3), or by first finetuning the Whisper model before adding the CTC head (as in M4). Comparing M4 and M6, we see that, training Whisper model alongside the CTC head degraded the WER slightly. However, we see in Section~\ref{sec:xcorpus} that M6 is more robust than M4 on conversational speech. Training only the encoder of Whisper alongside the CTC head (M2 and M5) degrades both the WER and the boundary error. From the table, we can also see that finetuning of the decoder, which tunes the implicit language model of the encoder-decoder models, is important while adapting Whisper to new datasets. Note that, it is possible to read out the transcript from the CTC head output. But our models were trained for a fixed number of epochs, without a criterion on ASR performance on CTC head. The WER from the CTC head output is typically high and hence it is not reported.

\vspace{-2pt}
\subsection{Cross-Corpus Generalisation}\label{sec:xcorpus}
\begin{table}[t]
\centering
\setlength{\tabcolsep}{4pt}
\caption{ASR performance on the four test tests. HW denotes the Hindi-Whisper model. 
}
\label{tab:xcorpfull}
\vspace{3pt}
\begin{tabular}{llrrrrr}
\toprule
Corpus & Model & WER & Sub & Del & Ins & CER \\
\midrule
\multirow{4}{*}{IndicVoices-R}
 & M3 & \textbf{14.38} & \textbf{12.16} & 1.27 & \textbf{0.95} & \textbf{6.79} \\
 & M4         & 14.64 & 12.47 & \textbf{1.16} & 1.00 & 7.03 \\
 & M6         & 14.90 & 12.53 & 1.22 & 1.16 & 7.15 \\
 & HW         & 22.89 & 16.86 & 5.44 & 0.58 & 13.36 \\
\midrule
\multirow{4}{*}{MPS}
 & M3 & 14.66 & 12.12 & 1.71 & \textbf{0.83} & 8.61 \\
 & M4         & \textbf{14.39} & 12.01 & 1.24 & 1.14 & \textbf{8.42} \\
 & M6         & 14.69 & 12.52 & 1.27 & 0.90 & 8.58 \\
  & HW         & 21.30 & 15.79 & 4.87 & 0.64 & 13.33 \\ 
\midrule
\multirow{4}{*}{GramVaani}
 & M3 & 24.98 & 18.42 & 3.01 & \textbf{3.56} & \textbf{16.55} \\
 & M4         & \textbf{24.86} & 17.95 & 3.13 & 3.77 & 16.70 \\
 & M6         & 26.60 & 19.90 & \textbf{2.83} & 3.87 & 17.55 \\
  & HW         & 26.22 & 18.70 & 4.94 & 2.57 & 18.72 \\ 
\midrule
\multirow{4}{*}{DISPLACE-M}
 & M3 & \textbf{35.60} & \textbf{21.34} & 6.98 & 7.28 & 27.82 \\
 & M4         & 39.10 & 22.63 & \textbf{4.75} & 11.73 & 31.46 \\
 & M6         & 36.03 & 24.15 & 5.04 & \textbf{6.83} & \textbf{26.98} \\
 & HW         & 67.34 & 15.36 & 25.57 & 26.41 & 62.61 \\
\bottomrule
\end{tabular}
\vspace{-15pt}
\end{table}

Table~\ref{tab:xcorpfull} shows the ASR performance of the proposed models with CTC head on the four test sets described in Section~\ref{sec:setup}. Comparing HW and M4, we see that mere finetuning of the base Whisper model on IndicVoices-R improves the performance on all the test sets. Addition of the CTC head (in M3, M6) does not alter the performance significantly on IndicVoices-R, MPS and GramVaani datasets. But the performance improves on DISPLACE-M, which has conversational speech. A closer look at the constituent errors shows that the improvement in WER is largely due to a reduction of insertion errors. Conversational speech often contains very short speech segments, ($\approx 30\%$ of segments in DISPLACE-M data are shorter than $1$~s), which are prone to hallucination. We infer that the added CTC head acts as a regularizer limiting the instances of hallucination, and hence reducing insertions and improving the overall WER. This can be seen across the test sets, the variants M3 and M6 have less number of insertions and smaller CER compared to M4 and HW.



\vspace{-2pt}
\subsection{ASR Error Correction}
\label{sec:apply}
A closer look at the distribution of DISPLACE-M word errors with the segment duration (results not shown here in the interest of space) has shown us that the $30\%$ samples with $<1$~s duration in the dataset had a WER of $121.4\%$ using the M3 model, which was largely due to the insertions in the transcripts of these samples (sentence completions, repeated generations, etc. \cite{koenecke2024careless}). Hence, we studied correction of the transcript, using the information extracted from the CTC head output. First, we note that a simple beam search during the decoding, often, deteriorates the transcription in such samples. 
We studied four policies, first two \emph{replace} the decoder transcript of a segment with the one read greedily off the CTC head: (i) \textit{conf.} fires when the mean CTC alignment confidence of the decoder's words falls below $0.2$, and (ii) \textit{rep.} fires when the decoder emits more than $3\times$ as many characters as the CTC head. The second two \emph{truncate} the decoder transcript to a maximum character length, cutting at a word boundary, and differ only in the selection of the length threshold: (iii) \textit{dur.} takes it from the audio, one character per $20$~ms encoder frame, and (iv) \textit{cap.} takes it from the CTC head, $1.5\times$ its character count. 

\begin{table}[t]
\centering
\setlength{\tabcolsep}{2pt}
\caption{Performance with ASR error correction on DISPLACE-M using our models. "Segments" row gives the number of utterances where the correction was applied. 
}
\label{tab:repair}
\vspace{3pt}
\begin{tabular}{lrrrrrrr}
\toprule
 & \multicolumn{2}{c}{Decoding} & \multicolumn{5}{c}{ASR error correction} \\
\cmidrule(lr){2-3}\cmidrule(lr){4-8}
 & greedy & beam-5 & conf. & rep. & dur. & cap. & oracle \\
\midrule
\multicolumn{8}{c}{\textit{M3 configuration}}\\
WER & 35.60 & 41.15 & 35.51 & 32.72 & 32.88 & {32.20} & 31.25 \\
Ins & 7.28 & 13.37 & 7.15 & 4.37 & 4.56 & {3.66} & 3.29 \\
Del & 6.98 & 7.39 & 7.05 & 7.05 & 6.98 & 8.08 & 6.65 \\
Segments & N.A & N.A & 409 & 356 & 201 & 1{,}975 & 1{,}400 \\
\midrule
\multicolumn{8}{c}{\textit{M4 configuration}}\\
WER\,(\%)            & 39.10 & 44.68   & 38.94   & 33.26   & 33.90   & {32.64}   & 31.69 \\
Ins.\,(\%)           & 11.73 & 17.93   & 11.46   & 5.77    & 6.52    & {4.79}    & 4.18 \\
Del.\,(\%)           &  4.75 &  4.75   &  4.84   & 4.83    & 4.75    & 6.19             & 4.97 \\
Segments             & N.A & N.A     & 609     & 624     & 377     & 2{,}795 & 1{,}361 \\
\midrule
\multicolumn{8}{c}{\textit{M6 configuration}}\\
 WER & 36.03 & 37.95 & 35.85 & 34.43 & 34.82 & {34.18} & 32.78 \\
 Ins & 6.83 & 9.20 & 6.62 & 5.21 & 5.62 & {4.81} & 3.99 \\
 Del & 5.04 & 5.06 & 5.11 & 5.11 & 5.04 & 5.89 & 5.11 \\
 Segments & N.A & N.A & 469 & 269 & 113 & 1{,}680 & 1{,}998 \\
\bottomrule
\end{tabular}
\vspace{-10pt}
\end{table}

\noindent Table~\ref{tab:repair} shows the ASR performance for different correction policies. We see that beam search increased the WER (largely due to increased in insertions) in all the configurations studied. We attribute this to the model's tendency to generate longer and complete sentences, the chance of which increases with beam search. Results in the column \textit{conf.} show that, the decoder token confidence cannot predict the hallucinations. The output token length in the \textit{rep.} policy is effective, but the policy discards the Whisper transcript and substitutes CTC policy for it, which is poor in general. The strategies, based on the truncation of Whisper output to a specific length (\textit{dur.}, \textit{cap.}), improve the WER and reduce the insertions, with a slight increase in the deletions. The \textit{cap} policy, where the transcript is truncated based on the character sequence length predicted by the CTC is head is more effective and reaches close to the upper limit indicated by \textit{oracle}. The trend is found to be similar across the three model configurations studied. 



\vspace{-3pt}
\section{Conclusion}
\vspace{-2pt}
We proposed a timestamped speech transcription system for Hindi, using an additional CTC head trained on the Whisper's encoder and forced alignment. Different training strategies were explored and the models were evaluated across four different datasets covering a range of domains. Our results show that the timestamp accuracy is comparable to that of WhisperX, which uses an external CTC trained model for force alignment. Further, our results show that finetuning Whisper's encoder and decoder alongside the added CTC head improves the generalization performance of Whisper and helps in reducing hallucinations. Simple thresholding strategies were explored for detecting segments with hallucinations, which proved effective in short speech segments. Future directions include an investigation of the variants of joint CTC/attention decoding for better transcription. 

\unnumberedsection{Acknowledgements}
We sincerely thank Professor Preeti Rao, Electrical Engineering, IIT Bombay for sharing the MPS-Hindi data and the useful discussions through the course of this project.
\unnumberedsection{Compliance with Ethical Standards}
IndicVoices-R and GramVaani are publicly available under their respective licences.
MPS-Hindi is not yet publicly released and is used here with the collectors' permission.  The
DISPLACE-M development set is distributed to participants of the Third DISPLACE Challenge
\cite{dhanya2026displacem} rather than released publicly, and is used here with the
organisers' consent; one author is a challenge organiser, which we disclose for
transparency.  The word-boundary annotations reported in
Sec.~\ref{sec:setup} were produced by two adult annotators who participated voluntarily and
with informed consent, annotating only already-public recordings.  No personal data was
collected and no new recordings were made, so ethical approval was not required.

\balance
\footnotesize
\bibliographystyle{IEEEbib}
\bibliography{refs}

@inproceedings{radford2023whisper,
  title={Robust speech recognition via large-scale weak supervision},
  author={Radford, Alec and Kim, Jong Wook and Xu, Tao and Brockman, Greg and
          McLeavey, Christine and Sutskever, Ilya},
  booktitle={Proc. Int. Conf. on Machine Learning (ICML)}, year={2023}}

@inproceedings{graves2006ctc,
  title={Connectionist temporal classification: Labelling unsegmented sequence data
         with recurrent neural networks},
  author={Graves, Alex and Fern{\'a}ndez, Santiago and Gomez, Faustino and
          Schmidhuber, J{\"u}rgen},
  booktitle={Proc. Int. Conf. on Machine Learning (ICML)}, pages={369--376}, year={2006}}

@article{watanabe2017hybrid,
  title={Hybrid {CTC}/attention architecture for end-to-end speech recognition},
  author={Watanabe, Shinji and Hori, Takaaki and Kim, Suyoun and Hershey, John R.
          and Hayashi, Tomoki},
  journal={IEEE J. Sel. Topics Signal Process.}, volume={11}, number={8},
  pages={1240--1253}, year={2017}}

@inproceedings{baevski2020wav2vec2,
  title={wav2vec 2.0: A framework for self-supervised learning of speech representations},
  author={Baevski, Alexei and Zhou, Yuhao and Mohamed, Abdelrahman and Auli, Michael},
  booktitle={Advances in Neural Information Processing Systems (NeurIPS)},
  volume={33}, pages={12449--12460}, year={2020}}

@inproceedings{conneau2021xlsr,
  title={Unsupervised cross-lingual representation learning for speech recognition},
  author={Conneau, Alexis and Baevski, Alexei and Collobert, Ronan and
          Mohamed, Abdelrahman and Auli, Michael},
  booktitle={Proc. Interspeech}, pages={2426--2430}, year={2021}}

@article{pratap2024mms,
  title={Scaling speech technology to 1000+ languages},
  author={Pratap, Vineel and Tjandra, Andros and Shi, Bowen and Tomasello, Paden and
          Babu, Arun and Kundu, Sayani and Elkahky, Ali and Ni, Zhaoheng and
          Vyas, Apoorv and Fazel-Zarandi, Maryam and others},
  journal={J. Machine Learning Research}, volume={25}, number={97}, pages={1--52},
  year={2024}}

@inproceedings{bain2023whisperx,
  title={{WhisperX}: Time-accurate speech transcription of long-form audio},
  author={Bain, Max and Huh, Jaesung and Han, Tengda and Zisserman, Andrew},
  booktitle={Proc. Interspeech}, pages={4489--4493}, year={2023}}

@inproceedings{wagner2024crisperwhisper,
  title={{CrisperWhisper}: Accurate timestamps on verbatim speech transcriptions},
  author={Wagner, Laurin and Thallinger, Bernhard and Zusag, Mario},
  booktitle={Proc. Interspeech}, year={2024}}

@misc{lintoai2023whispertimestamped,
  title={whisper-timestamped},
  author={Louradour, J{\'e}r{\^o}me},
  journal={GitHub repository},
  year={2023},
  publisher={GitHub},
  howpublished = {\url{https://github.com/linto-ai/whisper-timestamped}}
}

@misc{whisper-dtw,
  title={openai-dtw},
  author={J. W. Kim},
  journal={GitHub repository},
  year={2023},
  publisher={GitHub},
  howpublished = {\url{https://github.com/openai/whisper/blob/main/notebooks/Multilingual_ASR.ipynb}}
}

@inproceedings{mcauliffe2017mfa,
  title={{M}ontreal {F}orced {A}ligner: Trainable text-speech alignment using {K}aldi},
  author={McAuliffe, Michael and Socolof, Michaela and Mihuc, Sarah and
          Wagner, Michael and Sonderegger, Morgan},
  booktitle={Proc. Interspeech}, pages={498--502}, year={2017}}

@inproceedings{kurzinger2020ctcseg,
  title={{CTC}-segmentation of large corpora for {German} end-to-end speech recognition},
  author={K{\"u}rzinger, Ludwig and Winkelbauer, Dominik and Li, Lujun and
          Watzel, Tobias and Rigoll, Gerhard},
  booktitle={Proc. Int. Conf. on Speech and Computer (SPECOM)},
  pages={267--278}, year={2020}}

@inproceedings{zhou2025streamingwhisper,
  title={Adapting {W}hisper for streaming speech recognition via two-pass decoding},
  author={Zhou, Haoran and Song, Xingchen and Fahy, Brendan and Song, Qiaochu and
          Zhang, Binbin and Peng, Zhendong and Wadhawan, Anshul and Jiang, Denglin and
          Verma, Apurv and Ramesh, Vinay and Prasad, Srivas and
          Franceschini, Michele M.},
  booktitle={Proc. Interspeech}, year={2025}}

@article{polok2025dicow,
  title={{DiCoW}: Diarization-conditioned {W}hisper for target speaker
         automatic speech recognition},
  author={Polok, Alexander and Klement, Dominik and Kocour, Martin and Han, Jiangyu and
          Landini, Federico and Yusuf, Bolaji and Wiesner, Matthew and
          Khudanpur, Sanjeev and {\v{C}}ernock{\'y}, Jan and Burget, Luk{\'a}{\v{s}}},
  journal={Computer Speech \& Language}, year={2025}}

@inproceedings{koenecke2024careless,
  title={Careless {W}hisper: Speech-to-text hallucination harms},
  author={Koenecke, Allison and Choi, Anna Seo Gyeong and Mei, Katelyn X. and
          Schellmann, Hilke and Sloane, Mona},
  booktitle={Proc. ACM Conf. on Fairness, Accountability, and Transparency (FAccT)},
  pages={1672--1681}, year={2024}}

@inproceedings{indicvoicesr2024,
  title={{INDICVOICES-R}: Unlocking a massive multilingual multi-speaker speech corpus
         for scaling {Indian} {TTS}},
  author={Sankar, Ashwin and Anand, Srija and Varadhan, Praveen and Thomas, Sherry and
          Singal, Mehak and Kumar, Shridhar and Mehendale, Deovrat and
          Krishana, Aditi and Raju, Giri and Khapra, Mitesh},
  booktitle={Advances in Neural Information Processing Systems (NeurIPS),
             Datasets and Benchmarks Track}, year={2024}}

@inproceedings{bhanushali2022gramvaani,
  title={{G}ram {V}aani {ASR} challenge on spontaneous telephone speech recordings
         in regional variations of {H}indi},
  author={Bhanushali, Anish and Bridgman, Grant and Deekshitha, G. and
          Ghosh, Prasanta and Kumar, Pratik and Kumar, Saurabh and
          Kolladath, Adithya Raj and Ravi, Nithya and Seth, Aaditeshwar and
          Seth, Ashish and Singh, Abhayjeet and Sukhadia, Vrunda N. and
          Umesh, S. and Vyas, Sathvik and Yarra, Chiranjeevi},
  booktitle={Proc. Interspeech}, pages={3548--3552}, year={2022}}

@article{dhanya2026displacem,
  title={Benchmarking speech systems for frontline health conversations:
         The {DISPLACE-M} challenge},
  author={Dhanya, E and Meena, Ankita and Nanivadekar, Manas and Noumida, A and
          Azad, Victor and Shenoy, Ashwini Nagaraj and Chowdhuri, Pratik Roy and
          Banga, Shobhit and Chhabra, Vanshika and Bhat, Chitralekha and
          Kalluri, Shareef Babu and Chetupalli, Srikanth Raj and
          Vijayasenan, Deepu and Ganapathy, Sriram},
  journal={arXiv preprint arXiv:2603.02813}, year={2026}}

@inproceedings{loshchilov2019adamw,
  title={Decoupled weight decay regularization},
  author={Loshchilov, Ilya and Hutter, Frank},
  booktitle={Proc. Int. Conf. on Learning Representations (ICLR)}, year={2019}}

@inproceedings{loshchilov2017sgdr,
  title={{SGDR}: Stochastic gradient descent with warm restarts},
  author={Loshchilov, Ilya and Hutter, Frank},
  booktitle={Proc. Int. Conf. on Learning Representations (ICLR)}, year={2017}}

@misc{vanheeswijk2021jiwer,
  title={{jiwer}: Evaluate your speech-to-text system with similarity measures},
  author={Vaessen, Nik},
  year={2021}, publisher={GitHub},
  howpublished={\url{https://github.com/jitsi/jiwer}}}

@inproceedings{gothi2024miscue,
  title={A dataset and two-pass system for reading miscue detection},
  author={Gothi, Raj and Kumar, Rahul and Pereira, Mildred and Nayak, Nagesh and
          Rao, Preeti},
  booktitle={Proc. Interspeech}, year={2024}}

@inproceedings{raman2025oralreading,
  title={Oral reading errors by grade 3 children in {I}ndian schools:
         A {H}indi--{E}nglish perspective},
  author={Raman, Sneha and Rao, Preeti},
  booktitle={Proc. Interspeech}, year={2025}}

@misc{NemoFA,
  author       = {{NVIDIA}},
  title        = {{NeMo Forced Aligner}},
  howpublished = {\url{https://github.com/NVIDIA/NeMo/tree/main/tools/nemo_forced_aligner}},
  note         = {Part of the NVIDIA NeMo toolkit; accessed 2026},
  year         = {2024},
}

\end{document}